\documentstyle [12pt]{article}
\advance\textwidth by +1.0in
\advance\textheight by +1.0in
\advance\oddsidemargin by -0.5in
\advance\evensidemargin by -1.0in
\advance\topmargin by -0.5in
\def\ov#1{\overline{#1}}

\def\vb#1{\mbox{\boldmath$#1$}}
\def\pd#1#2{\frac{\partial #1}{\partial #2}}

\def\wh#1{\widehat{#1}}
\def\bdot{\,\vb{\cdot}\,}
\def\btimes{\,\vb{\times}\,}

\newcommand{\no}{\noindent}
\newcommand{\bc}{\begin{center}}
\newcommand{\ec}{\end{center}}

\begin{document}

\begin{flushright}
November 26, 2002
\end{flushright}

\begin{center}
{\sf Linear Wave Spectrum associated with} \\
{\sf Collective Neutrino-Plasma Interactions in the Early Universe} \\

\vspace*{0.3in}

Alain J.~Brizard and Sarah L.~McGregor \\
{\it Department of Chemistry and Physics, Saint Michael's College} \\
{\it Colchester, Vermont 05439} \\

\end{center}

\vspace*{0.3in}

The effects of collective neutrino-plasma interactions on the linear wave spectrum supported by a magnetized electron-positron plasma in the presence of a neutrino-antineutrino medium are investigated. When a pair-symmetric background neutrino-plasma medium is perturbed by space-charge waves (electrostatic waves associated with electron-positron charge separation), our analysis shows that the neutrino and antineutrino fluids also separate and the pair symmetry of the background medium is broken. The cosmological implications of this pair-symmetry breaking mechanism are briefly discussed.

\vspace*{0.3in}


\vfill\eject

\bc
{\sf I. INTRODUCTION}
\ec

\vspace*{0.2in}

\no
{\bf A. Neutrino Cosmology}

\vspace*{0.1in}

Neutrino Cosmology is one of the hottest topics in theoretical and experimental cosmology following the recent discovery that neutrinos may be massive \cite{nu_oscillations}. The cosmological implications of massive neutrinos, which include the impact of massive neutrinos 
on the cosmic microwave background radiation \cite{Hannestad,CNB_1} and big bang nucleosynthesis, have recently been reviewed by Dolgov \cite{Dolgov} and Raffelt 
\cite{Raffelt}. 

Neutrinos interact with matter through the charged and neutral weak currents associated with 
the weak interaction \cite{Taylor}, whose strength is expressed in terms of the 
weak-interaction coupling constant $G_{F}$ (with $G_{F}/\hbar^{3}c^{3} \simeq 1.16 \times 
10^{-5}\;{\rm GeV}^{-2}$). Charged weak currents involve the exchange of the charged vector bosons $W^{\pm}$ ($m_{W} \simeq 80$ GeV/c$^{2}$) associated with processes involving leptons $\sigma$ interacting with neutrinos $\nu_{\sigma}$ of the same flavor (or generation), while 
neutral weak currents involve the exchange of the neutral vector boson $Z^{0}$ ($m_{Z} \simeq 91$ GeV/c$^{2}$) associated with processes involving neutrinos of all types interacting with arbitrary charged and neutral particles. 

A measure of the strength of weak interactions is expressed in terms of the neutrino mean-free-path $\ell_{\nu} \simeq 1.4 \times 10^{11}\;T_{{\rm MeV}}^{-5}$ cm \cite{Kolb_Turner}, where $T_{{\rm MeV}}$ denotes a characteristic temperature expressed in MeV units. Although 
neutrino-matter interactions are very weak under normal conditions ($T \ll 1$ MeV), we note 
that the ratio of the neutrino-matter collision rate ($c/\ell_{\nu} \simeq 0.21\;
T_{{\rm MeV}}^{5}$ sec$^{-1}$) to the Hubble expansion rate $(cH \simeq 0.67\;T_{{\rm MeV}}^{2}$ sec$^{-1}$) is larger than unity for neutrino-matter temperatures above 1-2 MeV. Such high temperatures occured within the first second after the Big Bang \cite{Kolb_Turner} and, consequently, the neutrino medium is strongly coupled to the matter medium during the Early Universe ($T > 1$ MeV). 

Neutrino-matter interactions are described either in terms of discrete-particle collisional effects, with typical scattering cross-sections which scale with the strength of weak-interactions as $G_{F}^{2}$ (e.g., $\ell_{\nu} \propto G_{F}^{-2}$), or in terms of collective (self-consistent) effects, which scale as $G_{F}$ (see below). Although discrete-particle effects tend to dominate over collective effects at very large temperatures \cite{dissipative}, collective effects may be important during the later eras of the Early Universe. Indeed, the characteristic frequency associated with collective plasma oscillations $(\omega_{p} \simeq 2.7 \times 10^{20}\;T_{{\rm MeV}}^{3/2}$ rad/sec) is larger than the neutrino-matter collision rate $c/\ell_{\nu}$ by a factor $\omega_{p}\ell_{\nu}/c \simeq 1.3 \times 10^{21}\;
T_{{\rm MeV}}^{-7/2}$, which is larger than unity when the neutrino-matter temperatures are smaller than $10^{6}$ MeV. In addition, the ratio of the plasma frequency to the photon-matter collision rate $(c/\ell_{\gamma} \simeq 4.8 \times 10^{17}\;T_{{\rm MeV}}^{3}$ sec$^{-1}$) is expressed as $\omega_{p}\ell_{\gamma}/c \simeq 560 \;T_{{\rm MeV}}^{-3/2}$, which exceeds unity for temperatures less than 70 MeV. Hence, the short-time evolution of the primordial 
neutrino-plasma medium (in the temperature range 1 MeV $< T <$ 10 MeV) appears to be dominated by collective (collisionless) effects. 

In the present paper, we therefore investigate collective neutrino-plasma interactions in the Early Universe when the neutrino-plasma temperatures are between 1-2 MeV and 10 MeV. This temperature range corresponds to the later stage of the Lepton era \cite{Kolb_Turner}, when the Universe is predominantly populated with photons, electrons, positrons, and their associated neutrinos and antineutrinos.

\vspace*{0.2in}

\no
{\bf B. Effective Weak-Interaction Charge $G_{\sigma\nu}$}

\vspace*{0.1in}

As a neutrino propagates through a stationary (unpolarized) matter medium, it experiences a force derived from an effective potential \cite{Notzold_Raffelt,Kuo_Pantaleone} of the form $\sum_{\sigma}\;G_{\sigma\nu}\,n_{\sigma}$, where $n_{\sigma}$ denotes the matter density associated with particle species $\sigma$ and $G_{\sigma\nu}$ denotes the effective weak-interaction {\it charge} associated with $\sigma-\nu$ interactions. For example, the effective potential experienced by electron neutrinos (or antineutrinos) propagating in a stationary electron-positron medium is expressed in terms of the effective weak-interaction charge \cite{Notzold_Raffelt,Kuo_Pantaleone} 
\begin{equation}
G_{\sigma\nu} \;=\; \frac{G_{F}}{\sqrt{2}} \left[\; (\delta_{\sigma e} - \delta_{\sigma\ov{e}})\,(\delta_{\nu\nu_{e}} - \delta_{\nu\ov{\nu}_{e}})\; \left( 1 + 4\,\sin^{2}\theta_{W}\right) \;-\; \frac{16\langle E_{\sigma}\rangle
\,E_{\nu}}{3\,m_{W}^{2}c^{4}} \;\right],
\label{eq:G_sigmanu}
\end{equation}
where the first term combines the lowest-order ($\propto G_{F}$) contributions of charged and neutral weak currents [with $\sin^{2}\theta_{W} = 1 - (m_{W}/m_{Z})^{2} \simeq 0.22$], while the second term involves higher-order $(\propto G_{F}/m_{W}^{2}$) contributions, where $E_{\nu}$ denotes the neutrino (antineutrino) energy and $\langle E_{\sigma}\rangle$ denotes the average matter-particle energy for species $\sigma$. For an electron-positron plasma, we note that the effective potential becomes
\begin{equation} 
\sum_{\sigma = e,\ov{e}}\;G_{\sigma\nu}\,n_{\sigma} \;=\; \frac{G_{F}}{\sqrt{2}} \left[\; \pm \left( 1 + 4\,\sin^{2}\theta_{W}\right)\;\Delta n_{e} \;-\; \left. \left. \frac{16\,E_{\nu}}{3\,m_{W}^{2}c^{4}} \right( \langle E_{e}\rangle\;n_{e} + \langle E_{\ov{e}}\rangle\;n_{\ov{e}} \right) \;\right],
\label{eq:enu_potential}
\end{equation}
where $\Delta n_{e} = n_{e} - n_{\ov{e}}$ denotes the electron-positron charge-separation density and the $(\pm)$ signs in the first term refer to $\nu_{e}$ neutrinos $(+)$ or $\ov{\nu}_{e}$ antineutrinos $(-)$. Note that the first term in Eq.~(\ref{eq:enu_potential}) is antisymmetric to the charge-parity (CP) interchange ($e \leftrightarrow \ov{e}$ and $\nu \leftrightarrow \ov{\nu}$) while the second term is CP-symmetric (see Ref.~\cite{Notzold_Raffelt} for further discussion). 

The charge neutrality of the Early Universe during the Lepton Era, however, dictates that the charge-separation density $\Delta n_{e}$ be very small (i.e., the charge separation density $\Delta n_{e}$ is equal, by quasi-neutrality, to the proton density $n_{p} \simeq 10^{-10}
\,n_{e}$). The ratio of the CP-asymmetric term to the CP-symmetric term in Eq.~(\ref{eq:enu_potential}) is expressed approximately as $(\Delta n_{e}/10^{-10}\,n_{e})/
4\,T_{{\rm MeV}}^{2}$ and, hence, the CP-symmetric term dominates over the CP-asymmetric term for neutrino-plasma energies above 1 MeV, i.e., for equilibrium conditions of primordial electron-positron plasmas typically found during the Lepton Era of the Early Universe.

If the primordial electron-positron plasma is perturbed by space-charge waves (electrostatic waves associated with electron-positron charge separation), however, neutrinos can feel the influence of the effective CP-asymmetric potential. We henceforth ignore higher-order corrections in Eq.~(\ref{eq:G_sigmanu}) and investigate collective neutrino-plasma interactions in an electron-positron plasma perturbed by space-charge waves (for which $\Delta n_{e} \gg 10^{-10}\,n_{e}$). In this {\it space-charge} scenario, the effective weak-interaction charge $G_{\sigma\nu}$ is assumed to possess the CP-symmetry property
\begin{equation}
G_{\sigma\nu} \;=\; -\,G_{\ov{\sigma}\nu} \;=\; -\,G_{\sigma\ov{\nu}} \;=\;
G_{\ov{\sigma}\ov{\nu}},
\label{eq:identity_G}
\end{equation}
where $\sigma = e$ ($\ov{\sigma} = \ov{e}$) and $\nu = \nu_{e}$ ($\ov{\nu} = \ov{\nu}_{e}$), 
and $G_{e\nu} \simeq \sqrt{2}\;G_{F}$. In the present work, we also ignore neutrino collective 
self-interactions \cite{Kuo_Pantaleone} since we are interested in collective neutrino-plasma effects only.

\vspace*{0.2in}

\no
{\bf C. Collective Neutrino-Plasma Interactions}

\vspace*{0.1in}

Our discussion of collective neutrino-plasma effects begins with the problem of a single neutrino propagating in an unpolarized matter medium composed of charged and/or neutral particles. When a neutrino (of species $\nu$) propagates (with velocity ${\bf v}_{\nu}$) in a moving matter medium, it interacts by weak interaction with matter particles (of species $\sigma$) and behaves as if it were under the influence of an effective force \cite{BMW}
\begin{equation} 
\sum_{\sigma}\; G_{\sigma\nu} \left[\; -\;\left( \nabla n_{\sigma} + \frac{1}{c}\;
\pd{\vb{\Gamma}_{\sigma}}{t} \right) \;+\; \frac{{\bf v}_{\nu}}{c}\btimes\nabla\btimes
\vb{\Gamma}_{\sigma} \;\right],
\label{eq:force_nu}
\end{equation}
where $n_{\sigma}$ and $\vb{\Gamma}_{\sigma} = n_{\sigma}{\bf v}_{\sigma}/c$ denote the density and normalized particle flux of the matter fluid of species $\sigma$.

Similarly, when a matter particle (of species $\sigma$) propagates (with velocity 
${\bf v}_{\sigma}$) in a moving neutrino medium, it interacts by weak interaction with 
neutrinos and behaves as if it were under the influence of an effective force \cite{BMW}
\begin{equation} 
\sum_{\nu}\; G_{\sigma\nu} \left[\; -\;\left( \nabla n_{\nu} + \frac{1}{c}\;
\pd{\vb{\Gamma}_{\nu}}{t} \right) \;+\; \frac{{\bf v}_{\sigma}}{c}\btimes\nabla\btimes
\vb{\Gamma}_{\nu} \;\right],
\label{eq:force_sigma}
\end{equation}
where $n_{\nu}$ and $\vb{\Gamma}_{\nu} = n_{\nu}{\bf v}_{\nu}/c$ denote the density and normalized particle flux of the neutrino fluid of species $\nu$. The $\nu-\sigma$ symmetry of the two effective forces (\ref{eq:force_nu}) and (\ref{eq:force_sigma}) arises from the variational formulation of collective neutrino-plasma interactions introduced in Ref.~\cite{BMW}. 

The existence of effective forces (\ref{eq:force_nu})-(\ref{eq:force_sigma}) on a single neutrino-matter particle implies that neutrino-matter fluids are self-consistenly coupled through collective weak-interaction effects. Hence, for example, electromagnetic fields can influence neutrino-fluid dynamics in the presence of a charged-plasma medium just as if neutrinos carried an effective electrical charge \cite{Mendoca_etal,Nieves,Bhattacharya_etal}. The purpose of the present work is to investigate the effects of collective neutrino-plasma interactions on the linear wave spectrum supported by a magnetized electron-positron plasma in the presence of a neutrino-antineutrino medium. 

\vspace*{0.2in}

\no
{\bf D. Organization}

\vspace*{0.1in}

The remainder of this paper is organized as follows. In Sec.~II, the basic model for collective neutrino-plasma interactions in the presence of an electromagnetic field is derived on the basis of nonlinear fluid equations derived by variational principle by Brizard, Murayama, and Wurtele \cite{BMW}. In Sec.~III, the nonlinear fluid equations derived in Sec.~II are linearized about 
a stationary uniform neutrino-plasma medium. We also assume the background medium to be pair 
symmetric, i.e., the electron lepton number density $L_{e} = n_{e} - n_{\ov{e}}$ and the neutrino lepton number density $L_{\nu} = n_{\nu} - n_{\ov{\nu}}$ are both assumed to be zero 
at equilibrium (i.e., the chemical potential for each fluid species is zero). Since the primordial plasma is thought to be magnetized \cite{magnetic}, our analysis also includes the presence of a uniform magnetic field. The normal-mode analysis of the linearized fluid equations leads to a dispersion relation whose branches include the standard plasma branches associated with electromagnetic and electrostatic waves propagating in a magnetized electron-positron plasma (see Refs.~\cite{pair_1}-\cite{pair_4}), now modified by collective neutrino-plasma effects and a pure neutrino-sound branch associated with bulk neutrino-density propagating waves. 

In Sec.~IV, we investigate space-charge waves \cite{KT} associated with electrostatic modes driven by electron-positron charge separation. In summarizing our work in Sec.~V, we discuss the 
potential implications of neutrino-modified space-charge waves for Early Universe Cosmology, since they provide an effective CP-symmetry breaking mechanism leading to Leptogenesis.

\vspace*{0.2in}

\bc
{\sf II. FLUID MODEL FOR COLLECTIVE NEUTRINO-PLASMA INTERACTIONS}
\ec

\vspace*{0.2in}

The equations for the relativistic fluid dynamics associated with collective neutrino-plasma interactions in the presence of an electromagnetic field were recently derived from a variational principle by Brizard, Murayama, and Wurtele in Ref.~\cite{BMW}. Each fluid is represented by its proper density $\gamma_{s}^{-1}\,n_{s}$, its fluid velocity ${\bf v}_{s}$ 
and its normalized particle-flux density $\vb{\Gamma}_{s} = n_{s}\,{\bf v}_{s}/c$, its fluid pressure $p_{s}$ and its normalized fluid enthalpy $w_{s} = h_{s}/m_{s}c^{2}$ (where enthalpy $h_{s} = \mu_{s} + S_{s}T_{s}$ is defined in terms of the chemical potential $\mu_{s}$, entropy $S_{s}$, and temperature $T_{s}$). Here, the fluid label $s$ is either $s = e$ and $\ov{e}$ for electrons and positrons, respectively (with mass $m_{\ov{e}} = m_{e}$, electric charge $q_{\ov{e}} = e = -\,q_{e}$, and effective weak charge $G_{\ov{e}\nu} = -\,G_{e\nu}$), or $s = \nu$ and $\ov{\nu}$ for electron-neutrinos and electron-antineutrinos, respectively (with mass $m_{\ov{\nu}} = m_{\nu} \neq 0$ and effective weak charge $G_{e\ov{\nu}} = -\,G_{e\nu}$).

Each fluid species $s$ satisfies a continuity equation
\begin{equation}
\frac{1}{c}\;\pd{n_{s}}{t} \;+\; \nabla\bdot\vb{\Gamma}_{s} \;=\; 0.
\label{eq:continuity}
\end{equation}
The relativistic force equations for the electron-positron plasma fluids are \cite{BMW}
\begin{eqnarray}
\frac{d_{e}}{dt} \left[\; m_{e}\,(1 + w_{e})\;\gamma_{e}
{\bf v}_{e} \;\right] & = & -\,n_{e}^{-1}\,\nabla p_{e} \;-\; e\; \left( {\bf E} \;+\;
\frac{{\bf v}_{e}}{c}\btimes{\bf B} \right) \nonumber \\
 & + & G_{e\nu} \left[\; -\;\left( \nabla \Delta n_{\nu} + \frac{1}{c}\,
\pd{\Delta\vb{\Gamma}_{\nu}}{t} \right) \;+\; \frac{{\bf v}_{e}}{c} \btimes  \nabla\btimes\Delta\vb{\Gamma}_{\nu} \right],
\label{eq:mom_e} \\
 \nonumber \\
\frac{d_{\ov{e}}}{dt} \left[\; m_{e}\,(1 + w_{\ov{e}}) \;\gamma_{\ov{e}}{\bf v}_{\ov{e}} \;\right] & = & -\,n_{\ov{e}}^{-1}\,\nabla p_{\ov{e}} \;+\; e\; \left( {\bf E} \;+\; 
\frac{{\bf v}_{\ov{e}}}{c}\btimes{\bf B} \right) \nonumber \\
 & - & G_{e\nu} \left[\; -\;\left( \nabla \Delta n_{\nu} + \frac{1}{c}\,\pd{\Delta\vb{\Gamma}_{\nu}}{t} \right) \;+\; \frac{{\bf v}_{\ov{e}}}{c} \btimes  \nabla\btimes\Delta\vb{\Gamma}_{\nu} \right],
\label{eq:mom_ebar}
\end{eqnarray}
where $d_{s}/dt = (\partial_{t} + {\bf v}_{s}\bdot\nabla)$ denotes a total convective time derivative for fluid species $s$ and the asymmetric neutrino-flux four-vector $\Delta \Gamma_{\nu}^{\alpha} = (\Delta n_{\nu},\,\Delta\vb{\Gamma}_{\nu})$ is defined as  $\Delta \Gamma_{\nu}^{\alpha} \equiv \Gamma_{\nu_{e}}^{\alpha} - \Gamma_{\ov{\nu}_{e}}^{\alpha}$. The first two sets of terms on the right side of Eqs.~(\ref{eq:mom_e}) and (\ref{eq:mom_ebar}) represent the classical pressure-gradient and electromagnetic forces. The remaining terms are associated with CP-asymmetric neutrino-antineutrino forces, which arise as a result of collective neutrino-plasma interactions.

Next, the relativistic force equations for neutrino-antineutrino fluids are \cite{BMW}
\begin{eqnarray}
\frac{d_{\nu}}{dt} \left[\; m_{\nu}\,(1 + w_{\nu})\; \gamma_{\nu}{\bf v}_{\nu} \;\right] & = & 
-\,n_{\nu}^{-1}\,\nabla p_{\nu} \;-\; G_{e\nu} \left( \nabla \Delta n_{e} \;+\; \frac{1}{c}\,\pd{\Delta \vb{\Gamma}_{e}}{t} \right) \nonumber \\
 &  &\mbox{}+\; \left. \left. \frac{{\bf v}_{\nu}}{c}\btimes \right( G_{e\nu}\;\nabla \btimes \Delta\vb{\Gamma}_{e} \right),
\label{eq:mom_nu} \\
\nonumber \\
\frac{d_{\ov{\nu}}}{dt} \left[\; m_{\nu}\, (1 + w_{\ov{\nu}})\; \gamma_{\ov{\nu}}
{\bf v}_{\ov{\nu}} \;\right] & = & -\,n_{\ov{\nu}}^{-1}\,\nabla p_{\ov{\nu}} \;+\; G_{e\nu} \left( \nabla \Delta n_{e} \;+\; \frac{1}{c}\,\pd{\Delta \vb{\Gamma}_{e}}{t} \right) 
\nonumber \\
 &  &\mbox{}-\; \left. \left. \frac{{\bf v}_{\ov{\nu}}}{c}\btimes \right( G_{e\nu}\;\nabla \btimes \Delta\vb{\Gamma}_{e} \right),
\label{eq:mom_nubar}
\end{eqnarray}
where the asymmetric particle-flux four-vector $\Delta \Gamma_{e}^{\alpha} = (\Delta n_{e},\,
\Delta\vb{\Gamma}_{e})$ is defined as $\Delta \Gamma_{e}^{\alpha} \equiv \Gamma_{e}^{\alpha} - \Gamma_{\ov{e}}^{\alpha}$. The driving forces appearing on the right side of
Eqs.~(\ref{eq:mom_nu}) and (\ref{eq:mom_nubar}) include not only the classical 
pressure-gradient forces but also the asymmetric electron-positron driving forces.

Lastly, the evolution of the electromagnetic field is expressed in terms of the Maxwell equations
\begin{equation}
\left. \begin{array}{rcl}
\nabla\bdot{\bf E} & = & -\,4\pi\,e\;\Delta n_{e} \\
\nabla\bdot{\bf B} & = & 0 \\
 &&  \\
\nabla\btimes{\bf E} & = & -\,c^{-1}\,\partial_{t}{\bf B} \\
\nabla\btimes{\bf B} & = & c^{-1}\,\partial_{t}{\bf E} \;-\; 4\pi\,e\;\Delta\vb{\Gamma}_{e}
\end{array} \right\},
\label{eq:Maxwell}
\end{equation}
which couple the evolution of the electromagnetic field to plasma charge densities and currents through the asymmetric electron-positron particle-flux four-vector. Note that using Eq.~(\ref{eq:Maxwell}) we find 
\begin{equation} 
\nabla \Delta n_{e} \;+\; \frac{1}{c}\,\pd{\Delta \vb{\Gamma}_{e}}{t} \;=\; -\;
\frac{\Box^{2}{\bf E}}{4\pi\,e} \;\;\;{\rm and}\;\;\; \nabla \btimes \Delta\vb{\Gamma}_{e} \;=\; 
\frac{\Box^{2}{\bf B}}{4\pi\,e},
\label{eq:dAlembertian}
\end{equation}
where $\Box^{2} = \nabla^{2} - c^{-2}\,\partial_{t}^{2}$ is the D'Alembertian operator. Thus, substituting these expressions into Eqs.~(\ref{eq:mom_nu}) and (\ref{eq:mom_nubar}), we find that collective neutrino-plasma interactions allow electromagnetic fields to influence neutrino dynamics just as if neutrinos carried an effective electric charge \cite{Mendoca_etal,Nieves,Bhattacharya_etal}.

In summary, the coupled equations (\ref{eq:mom_e})-(\ref{eq:Maxwell}) provide a self-consistent model for the collective neutrino-plasma interactions in the presence of an electromagnetic field. This self-consistency implies, for example, that electromagnetic fields can generate asymmetric neutrino flows in the presence of a plasma medium or that collective neutrino-plasma interactions can assist in generating electromagnetic fields \cite{BMW}. 

\vspace*{0.2in}

\bc
{\sf III. LINEAR WAVE SPECTRUM IN A MAGNETIZED PLASMA-NEUTRINO MEDIUM}
\ec

\vspace*{0.2in}

\no
{\bf A. Normal-mode Representation}

\vspace*{0.1in}

We now consider the linear wave spectrum supported by a warm electron-positron plasma (with densities $n_{\ov{e}0} = n_{e0}$ and temperatures $T_{\ov{e}0} = T_{e0}$) interacting with a warm neutrino-antineutrino fluid (with densities $n_{\ov{\nu}0} = n_{\nu 0}$ and temperatures $T_{\ov{\nu}0} = T_{\nu 0}$) in the presence of a uniform magnetic field ${\bf B}_{0} = B_{0}
\,\wh{z}$. 

We first introduce the following representation for the electromagnetic field
\begin{equation}
\left. \begin{array}{rcl}
{\bf E} & = & \delta{\bf E} \\
{\bf B} & = & {\bf B}_{0} \;+\; \delta{\bf B}
\end{array} \right\},
\label{eq:pert_field}
\end{equation}
where we take the background magnetic field to be ${\bf B}_{0} = B_{0}\,\wh{{\sf z}}$, and the neutrino-plasma medium
\begin{equation}
\left. \begin{array}{rcl}
n_{s} & = & n_{s 0} \;+\; \delta n_{s} \\
\vb{\Gamma}_{s} & = & \delta \vb{\Gamma}_{s} \;=\; n_{s0}\,\delta{\bf v}_{s}/c \\
p_{s} & = & p_{s 0} \;+\; (dp_{s 0}/dn_{s 0})\;\delta n_{s} 
\end{array} \right\},
\label{eq:pert_enu}
\end{equation}
where we assume (for simplicity) that the scalar pressures $p_{e0}$ and $p_{\nu 0}$ satisfy the equation of state $p_{s0} = p_{s0}(n_{s0})$. We adopt a semi-relativistic approach based on the expansion of the relativistic factors $\gamma_{s}$ to second order in $|{\bf v}_{s}|^{2}/c^{2}$ while retaining the enthalpy correction to the particle's mass $M_{s} = m_{s} + h_{s0}/c^{2}$. We also introduce the dimensionless quantity 
\[ K_{s} \;=\; \frac{1}{M_{s}c^{2}}\;\frac{dp_{s 0}}{d n_{s 0}} \] 
defined as the square of the ratio of the speed of sound in fluid species $s$ to the speed of light.

Next, we introduce the normal-mode representation for the perturbation fields given in Eqs.~(\ref{eq:pert_field}) and
(\ref{eq:pert_enu}), where $\delta(\cdots) \rightarrow \delta(\cdots)\,\exp(i{\bf k}\bdot{\bf x} - i\omega t)$ and 
${\bf k} = k\,(\cos\theta\,\wh{{\sf z}} + \sin\theta\,\wh{{\sf x}}) \equiv k\,\wh{{\sf k}}$. We also introduce the bulk energy density and (normalized) bulk energy density flux
\begin{equation}
\left. \begin{array}{rcl} 
\delta\varrho_{s} & = & 4\pi M_{s}c^{2}\;(\delta n_{s} + \delta n_{\ov{s}}) \\
\\
\delta{\bf p}_{s} & = & 4\pi M_{s}c^{2}\;(\delta\vb{\Gamma}_{s} + \delta\vb{\Gamma}_{\ov{s}})
\end{array} \right\},
\label{eq:ep_densities}
\end{equation}
and the CP-asymmetry density and (normalized) current
\begin{equation}
\left. \begin{array}{rcl} 
\delta Q_{s} & = & -\;(4\pi\,iec/\omega)\;\left( \delta n_{s} - \delta n_{\ov{s}} \right) \\
\\
\delta {\bf J}_{s} & = & -\;(4\pi\,iec/\omega)\;\left( \delta\vb{\Gamma}_{s} - \delta \vb{\Gamma}_{\ov{s}} \right)
\end{array} \right\}.
\label{eq:CP_densities}
\end{equation}
The normalization (\ref{eq:CP_densities}) is chosen so that the perturbed densities $\delta Q_{s}$ and $\delta{\bf J}_{s}$ have the same units as the perturbed electromagnetic fields.

\vspace*{0.2in}

\no
{\bf B. Linearized Coupled Wave Equations}

\vspace*{0.1in}

The normal-mode analysis involves 22 perturbation fields associated with the electron-positron fluid $(\delta\varrho_{e},\delta{\bf p}_{e};\,\delta Q_{e}, \delta{\bf J}_{e})$, the 
neutrino-antineutrino fluid $(\delta\varrho_{\nu},\delta{\bf p}_{\nu};\,\delta Q_{\nu}, \delta{\bf J}_{\nu})$, and the electromagnetic field $(\delta{\bf E}, \delta{\bf B})$. The continuity equations (\ref{eq:continuity}) for the electron-positron and neutrino-antineutrino fluids become
\begin{equation}
\omega\,\delta\varrho_{s} \;=\; {\bf k}c\bdot\delta{\bf p}_{s} \;\;\;{\rm and}\;\;\;
\omega\,\delta Q_{s} \;=\; {\bf k}c\bdot\delta{\bf J}_{s},
\label{eq:wk_continuity}
\end{equation}
where $s = e$ or $\nu$. The perturbed electromagnetic field $(\delta{\bf E}, \delta{\bf B})$ satisfies the linearized Maxwell equations
\begin{equation}
\left. \begin{array}{rcl}
{\bf k}c\bdot\delta{\bf E} & = & -\,\omega\;\delta Q_{e} \\
{\bf k}c\bdot\delta{\bf B} & = & 0 \\
 &&  \\
{\bf k}c\btimes\delta{\bf E} & = & \omega\;\delta{\bf B} \\
{\bf k}c\btimes\delta{\bf B} & = & -\,\omega\;\left( \delta{\bf E} \;+\; \delta{\bf J}_{e} \right)
\end{array} \right\}.
\label{eq:wk_Maxwell}
\end{equation}
We can combine the last two Maxwell equations to eliminate the perturbed magnetic field $\delta{\bf B}$ in favor of the perturbed electric field $\delta{\bf E}$ to obtain
\begin{equation} 
\left[\; (1 - N)\,{\bf I} \;+\; {\sf N} \;\right]\bdot\delta{\bf E} \;=\; -\;
\delta{\bf J}_{e},
\label{eq:lin_Max}
\end{equation}
where ${\sf N} = N\,\wh{{\sf k}}\wh{{\sf k}}$ and $N = k^{2}c^{2}/\omega^{2}$ denotes the square of the refractive index.

Next, the linearized versions of the electron and positron fluid equations (\ref{eq:mom_e}) and (\ref{eq:mom_ebar}) yield
\begin{eqnarray}
\left( {\bf I} \;-\; K_{e}\,{\sf N} \right)\bdot\delta{\bf p}_{e} & = & \delta{\bf J}_{e}\btimes{\bf B}_{0},
\label{eq:delta_pe} \\
\left( {\bf I} \;-\; K_{e}\,{\sf N} \right)\bdot\delta{\bf J}_{e} & = & -\; Y\;\delta{\bf E}
\;-\; X\;\delta{\bf p}_{e}\btimes\frac{{\bf B}_{0}}{B_{0}^{2}} \;-\; \alpha_{e} \left( {\bf I} \;-\; {\sf N} \right)\bdot \delta{\bf J}_{\nu},
\label{eq:delta_Je}
\end{eqnarray}
where we have substituted $({\bf k}c/\omega)\delta\varrho_{e} = {\sf N}\bdot\delta{\bf p}_{e}$ and we have defined $X = \omega_{c}^{2}/\omega^{2}$ and $Y = \omega_{p}^{2}/\omega^{2}$. Here, we have introduced the following enthalpy-modified neutrino-plasma parameters: the 
electron-positron plasma frequency 
\[ \omega_{p} \;\equiv\; \sqrt{\sum_{\sigma}\;\frac{4\pi\;n_{\sigma 0}
q_{\sigma}^{2}}{M_{\sigma}}} \;=\; \sqrt{\,\frac{8\pi\;n_{e0}e^{2}}{M_{e}}}, \]
the electron-positron gyrofrequency $\omega_{c} \equiv eB_{0}/M_{e}c$, and the dimensionless neutrino-plasma interaction coefficient 
\[ \alpha_{e} \;\equiv\; \frac{2\,n_{e 0}G_{e\nu}}{M_{e}c^{2}} \;\simeq\; 8 \times 10^{-12}
\;T_{{\em MeV}}^{3}\;\left( \frac{m_{e}}{M_{e}} \right), \] 
so that $\alpha_{e} \ll 1$ for $T_{e} \ll 10$ GeV. 

Similarly, the electron-neutrino and electron-antineutrino fluid equations (\ref{eq:mom_nu}) and (\ref{eq:mom_nubar}) 
become
\begin{eqnarray}
\left( {\bf I} \;-\; K_{\nu}\,{\sf N} \right)\bdot\delta{\bf p}_{\nu} & = & 0,
\label{eq:delta_pnu} \\
\left( {\bf I} \;-\; K_{\nu}\,{\sf N} \right)\bdot\delta{\bf J}_{\nu} & = & -\; \alpha_{\nu} \left( 
{\bf I} \;-\; {\sf N} \right)\bdot \delta{\bf J}_{e},
\label{eq:delta_Jnu}
\end{eqnarray}
where $\alpha_{\nu} \equiv 2\;n_{\nu 0}G_{e\nu}/M_{\nu}c^{2} \ll 1$. The corresponding equations 
for $\delta\varrho_{\nu}$ and $\delta Q_{\nu}$ can be obtained from Eqs.~(\ref{eq:delta_pnu}) and (\ref{eq:delta_Jnu}), respectively, by taking their dot product with $({\bf k}c/\omega)$ 
and use the continuity equations (\ref{eq:wk_continuity}) to obtain
\begin{eqnarray}
\left( 1 \;-\; K_{\nu}\,N\right)\;\delta\varrho_{\nu} & = & 0,
\label{eq:delta_rhonu} \\
\left( 1 \;-\; K_{\nu}\,N\right)\;\delta Q_{\nu} & = & -\;\alpha_{\nu}\;( 1 - N)\;\delta Q_{e}.
\label{eq:delta_Qnu}
\end{eqnarray}
We note that the neutrino-bulk equation (\ref{eq:delta_rhonu}) yields the linear dispersion relation $1 = K_{\nu}\,N$ or
$\omega^{2} \;=\; K_{\nu}\;k^{2}c^{2}$, which describes neutrino {\it sound} waves in a neutrino-antineutrino medium when $\delta\varrho_{\nu} \neq 0$ (i.e., when $\delta n_{\nu_{e}} \neq -\,\delta n_{\ov{\nu}_{e}}$). Under the model considered here (which retains only a CP-asymmetric weak-interaction charge), however, neutrino-sound waves are decoupled from collective neutrino-plasma effects.
 
\vspace*{0.2in}

\no
{\bf C. Linear Dispersion Relation for Collective Neutrino-Plasma Interactions}

\vspace*{0.1in}

We now show that collective neutrino-plasma interactions couple density perturbations $\delta n_{\nu_{e}}$ and $\delta n_{\ov{\nu}_{e}}$ in such a way that, if $\delta n_{\nu_{e}} \neq 0$, then $\delta\varrho_{\nu} = 0$ and $\delta Q_{\nu} \neq 0$, i.e., collective neutrino-plasma interactions generate CP asymmetries in the neutrino-antineutrino medium.

First, Eq.~(\ref{eq:delta_pe}) can be inverted to give
\begin{equation}
\delta{\bf p}_{e} \;=\; \left( {\bf I} \;+\; \frac{K_{e}}{\Delta_{e}}\,{\sf N} \right)\bdot
\delta{\bf J}_{e}\btimes{\bf B}_{0},
\label{eq:p_e}
\end{equation}
where $\Delta_{e} = 1 - K_{e}\,N = {\rm det}({\bf I} - K_{e}\,{\sf N})$. Next, Eq.~(\ref{eq:delta_Jnu}) can be inverted to yield
\begin{equation}
\delta{\bf J}_{\nu} \;=\; -\;\alpha_{\nu}\;\left[ {\bf I} \;+\; \frac{(K_{\nu} - 1)}{\Delta_{\nu}}\,{\sf N} \right]\bdot\delta{\bf J}_{e},
\label{eq:J_nu}
\end{equation}
where $\Delta_{\nu} = 1 - K_{\nu}\,N = {\rm det}({\bf I} - K_{\nu}\,{\sf N}) \neq 0$. Note that, according to Eq.~(\ref{eq:delta_Qnu}), the CP-asymmetry of the neutrino-antineutrino medium ($\delta Q_{\nu} \neq 0$) and charge separation $\delta Q_{e} \neq 0$ impliy that $1 - K_{\nu}\,N \neq 0$ and thus the neutrino-antineutrino bulk density must vanish $\delta\varrho_{\nu} = 0$ (or $\delta n_{\ov{\nu}_{e}} = -\,\delta n_{\nu_{e}}$).

Lastly, the second term on the right side of Eq.~(\ref{eq:delta_Je}) can be rewritten using Eq.~(\ref{eq:p_e}) as
\[ X\; \left[\; \wh{z}\btimes\left( {\bf I} \;+\; \frac{K_{e}}{\Delta_{e}}\;{\sf N} \right)
\btimes\wh{z} \;\right]\bdot\delta{\bf J}_{e}, \]
while the third term on the right side of Eq.~(\ref{eq:delta_Je}) can be rewritten using Eq.~(\ref{eq:J_nu}) as
\[ \beta\,\left[\; {\bf I} \;+\; (K_{\nu} + N - 2)\;\frac{{\sf N}}{\Delta_{\nu}}\;\right]\bdot\delta{\bf J}_{e}. \]
By combining these results and introducing the dielectric tensor $\vb{\epsilon}$ as
\begin{equation} 
\delta{\bf J}_{e} \;=\; \left( \vb{\epsilon} \;-\; {\bf I} \right)\bdot\delta{\bf E},
\label{eq:epsilon_def}
\end{equation}
we find the following non-vanishing components for the dielectric tensor
\begin{eqnarray*}
\epsilon_{xx} & = & 1 \;-\; \frac{Y}{{\cal D}}\; \left( 1 - \beta - {\cal R}N\,\cos^{2}\theta\right), \\
 &  & \\
\epsilon_{yy} & = & 1 \;-\; \frac{Y\,\Delta_{e}}{(1 - \beta)\,\Delta_{e} - X(1 - K_{e}N\,\cos^{2}\theta)}, \\
 &  & \\
\epsilon_{zz} & = & 1 \;-\; \frac{Y}{{\cal D}}\; \left( 1 - \beta - X - {\cal R}N\,\sin^{2}\theta\right), \\
 &  & \\
\epsilon_{xz} & = & \epsilon_{zx} \;=\; \frac{{\cal R}N}{{\cal D}}\;\cos\theta\sin\theta,
\end{eqnarray*}
where ${\cal R} = K_{e} + \beta\,(K_{\nu} + N - 2)/\Delta_{\nu}$ and
\[ {\cal D} \;=\; (1 - \beta)\,(1 - \beta - X - {\cal R}N) \;+\; X\,{\cal R}N\,\cos^{2}\theta. \]

By substituting Eq.~(\ref{eq:epsilon_def}) into Eq.~(\ref{eq:lin_Max}), the Maxwell equations become
\begin{equation} 
\left[\; \vb{\epsilon} \;-\; \left( N\;{\bf I} \;-\; {\sf N} \right) \;\right]
\bdot\delta {\bf E} \;\equiv\; {\sf D}(\omega,N,\theta)\bdot\delta{\bf E} \;=\; 0,
\label{eq:dis_eq}
\end{equation}
where the dispersion tensor ${\sf D}(\omega,N,\theta)$ has the non-vanishing components
\begin{equation}
\left. \begin{array}{rcl}
D_{xx} & = & \epsilon_{xx} \;-\; N\;\cos^{2}\theta \\
D_{yy} & = & \epsilon_{yy} \;-\; N \\
D_{zz} & = & \epsilon_{zz} \;-\; N\;\sin^{2}\theta \\
D_{xz} & = & \epsilon_{xz} \;+\; N\;\sin\theta\,\cos\theta \\
D_{zx} & = & \epsilon_{zx} \;+\; N\;\sin\theta\,\cos\theta
\end{array} \right\}.
\label{eq:D_components}
\end{equation}
Nontrivial solutions of Eq.~(\ref{eq:dis_eq}) exist for $\delta{\bf E}$ only if 
\begin{equation} 
0 \;=\; {\rm det}\;{\sf D}(\omega,N,\theta) \;=\; D_{yy}\; \left( D_{xx}\,D_{zz} \;-\; D_{xz}\,D_{zx} \right),
\label{eq:det_D}
\end{equation}
which holds if $D_{yy} = 0$ or $D_{xx}\,D_{zz} = D_{xz}\,D_{zx}$.  Note that the non-diagonal components $D_{xz}$ and $D_{zx}$ vanish at $\theta = 0$ (parallel propagation, $\wh{{\sf k}}
\;\|\;{\bf B}_{0}$) and $\theta = \pi/2$ (perpendicular propagation, $\wh{{\sf k}}\;\bot\;
{\bf B}_{0}$).

\vspace*{0.2in}

\no
{\bf D. Wave Polarizations and Dispersion Relations}

\vspace*{0.1in}

\no
{\it 1. Electromagnetic modes} 

\vspace*{0.1in}

Electromagnetic modes satisfy the condition ${\bf k}\bdot\delta{\bf E} = 0$, with $\delta{\bf B} = ({\bf k}c/\omega)
\btimes\delta{\bf E}$ and $\vb{\epsilon}\bdot\delta{\bf E} = N\,\delta{\bf E}$. The perturbed electron-positron fluid is represented by $\delta Q_{e} = 0$ (since electromagnetic waves preserve charge neutrality) and $\delta{\bf J}_{e} = -\,
(1 - N)\,\delta{\bf E}$ while the perturbed neutrino-antineutrino fluid is represented by $\delta Q_{\nu} = 0$ and $\delta
{\bf J}_{\nu} = -\,\alpha_{\nu}\,\delta{\bf J}_{e}$. Since the charge neutrality of the electron-positron plasma 
invalidates the assumptions of our space-charge model, we turn our attention to electrostatic modes.

\vspace*{0.1in}

\no
{\it 2. Electrostatic modes} 

\vspace*{0.1in}

Electrostatic modes satisfy the condition ${\bf k}\bdot\delta{\bf E} \neq 0$ and $\delta{\bf B} = 0$; these modes are also refered to as space-charge waves since $\delta Q_{e} = -\;({\bf k}
c/\omega)\bdot\delta{\bf E} \neq 0$ and $\delta{\bf J}_{e} = -\,\delta{\bf E}$, according to Eq.~(\ref{eq:wk_Maxwell}). We also note that space-charge waves can induced bulk motion in the electron-positron medium, with $\delta{\bf p}_{e} = -\,\delta{\bf E}\btimes{\bf B}_{0}$ and $\delta\varrho_{e} = 0$.

For parallel propagation ($\theta = 0$ and $\wh{{\sf k}} = \wh{{\sf z}}$), the wave polarization is $\delta{\bf E} = \delta E\,\wh{z}$ and the dispersion relation is the dispersion relation 
\begin {equation}
D_{zz} \;=\; \epsilon_{zz} \;=\; 1 \;-\; \left( \frac{Y}{1 - \beta - {\cal R}N} \right) \;=\; 0.  
\label{eq:D_zz}
\end{equation}
Here, collective neutrino-plasma effects (proportional to $\beta$) modify the plasma dispersion relation $K_{e}\,N = 1 - Y \equiv P$ or $\omega^{2} = \omega_{p}^{2} + k^{2}\,(K_{e}c^{2})$, where $K_{e}c^{2}$ represents the square of the speed of sound in the electron-positron medium (which includes effects due to finite enthalpy).

For perpendicular propagation ($\theta = \pi/2$ and $\wh{{\sf k}} = \wh{{\sf x}}$), the wave polarization is $\delta{\bf E} = \delta E\,\wh{x}$ and the dispersion relation is
\begin {equation}
D_{xx} \;=\; \epsilon_{xx} \;=\; 1 \;-\; \left( \frac{Y}{1 - \beta - X - {\cal R}N} \right) \;=\; 0.
\label{eq:D_xx}
\end{equation}
Here, collective neutrino-plasma effects (proportional to $\beta$) modify the plasma dispersion relation $K_{e}\,N = 1 - (X + Y) \equiv S$ or $\omega^{2} = (\omega_{p}^{2} + \omega_{c}^{2}) + k^{2}\,(K_{e}c^{2})$. 

In the next Section, we focus on electrostatic modes since space-charge waves (with $\delta Q_{e} \neq 0$) are coupled to CP-symmetry breaking effects $(\delta Q_{\nu} \neq 0)$.

\vspace*{0.2in}

\bc
{\sf IV. SPACE-CHARGE WAVES}
\ec

\vspace*{0.2in}

We now consider how space-charge (electrostatic) waves in a warm electron-positron plasma are affected by collective neutrino-plasma interactions (see Refs.~\cite{pair_1,pair_3} for further deatils). Since $\delta Q_{e} \neq 0$ for space-charge waves, we find
\begin{equation} 
\delta Q_{\nu} \;=\; -\;\frac{\alpha_{\nu}\,(1 - N)\;\delta Q_{e}}{(1 - K_{\nu}\,N)} \neq 0
\label{eq:Qnu_e}
\end{equation}
from (\ref{eq:delta_Qnu}). There are two branches associated with space-charge waves: the {\bf Langmuir} branch $D_{zz} = 0$ (for $\theta = 0$ and $\delta{\bf E}\;\|\;{\bf B}_{0}$) and the {\bf upper-hybrid} branch $D_{xx} = 0$ (for $\theta = \pi/2$ and $\delta{\bf E}\;\bot\;
{\bf B}_{0}$). 

\vspace*{0.2in}

\no
{\bf A. Langmuir Branch}

\vspace*{0.1in}

The dispersion relation (\ref{eq:D_zz}) for the Langmuir branch yields the quadratic equation
\begin{equation} 
(K_{e}K_{\nu} - \beta)\;N^{2} \;-\; (K_{e} + K_{\nu}\,P - 2\beta)\;N \;-\; (P - \beta) \;=\; 0,
\label{eq:P}
\end{equation}
where $P = 1 - Y$. There are two solutions $N_{\pm}$ to this equation:
\begin{equation}
N_{\pm} \;=\; \frac{(K_{e} + K_{\nu}\,P - 2\beta) \;\pm\; \left[ (K_{e} - K_{\nu}\,P)^{2} \;+\;
4\beta\;(P - K_{e})\,(1 - K_{\nu})\right]^{1/2}}{2\;(K_{e}K_{\nu} - \beta)}. 
\label{eq:N_pm}
\end{equation}
Expanding the solution $N_{+}$ up to first order in $\beta$ (for $K_{s} \neq 0$) yields
\begin{equation}
N_{+} \;=\; \frac{1}{K_{\nu}} \left[\; 1 \;+\; \beta\; \frac{(1 - K_{\nu})^{2}}{K_{\nu}\,
(K_{e} - K_{\nu}\,P)} \;+\; \cdots \;\right],
\label{eq:NP_plus}
\end{equation}
so that
\[ 1 \;-\; N_{+} \;\simeq\; - \left( \frac{1}{K_{\nu}} - 1\right) \;\;\;{\rm and}\;\;\;
1 \;-\; K_{\nu}\,N_{+} \;\simeq\; -\; \frac{\beta\;(1 - K_{\nu})^{2}}{K_{\nu}\,
(K_{e} - K_{\nu}\,P)}. \]
Substituting these expressions into Eq.~(\ref{eq:Qnu_e}) yields
\begin{equation} 
\delta Q_{\nu}^{(+)}(P) \;\simeq\; -\; \left( \frac{K_{e} - K_{\nu}P}{1 - K_{\nu}} \right) 
\frac{\delta Q_{e}}{\alpha_{e}}.
\label{eq:Langmuir_plus}
\end{equation}
Since the neutrino-sound waves are recovered from the solution for $N_{+}$ in the absence of neutrino-plasma coupling $(\beta = 0)$, we call this solution the {\it neutrino} solution of 
the Langmuir branch.

Similarly, expanding the solution $N_{-}$ up to first order in $\beta$ (for $K_{s} \neq 0$) yields
\begin{equation}
N_{-} \;=\; \frac{1}{K_{e}} \left[\; P \;-\; \beta\; \frac{(K_{e} - P)^{2}}{K_{e}\,
(K_{e} - K_{\nu}\,P)} \;+\; \cdots \;\right],
\label{eq:NP_minus}
\end{equation}
so that
\[ 1 \;-\; N_{-} \;\simeq\; - \left( \frac{P}{K_{e}} - 1\right) \;\;\;{\rm and}\;\;\;
1 \;-\; K_{\nu}\,N_{-} \;\simeq\; \left( 1 \;-\; \frac{K_{\nu}}{K_{e}}\;P \right). \]
Substituting these expressions into Eq.~(\ref{eq:Qnu_e}) now yields
\begin{equation} 
\delta Q_{\nu}^{(-)}(P) \;\simeq\; \alpha_{\nu} \left( \frac{P - K_{e}}{K_{e} - K_{\nu}\,P} \right) \delta Q_{e}.
\label{eq:Langmuir_minus}
\end{equation}
Since the standard Langmuir plasma waves are recovered from the solution for $N_{-}$ in the absence of neutrino-plasma coupling $(\beta = 0)$, we call this solution the {\it plasma} solution of the Langmuir branch.

\vspace*{0.2in}

\no
{\bf B. Upper-Hybrid Branch}

\vspace*{0.1in}

A similar treatment of the upper-hybrid branch is facilitated by the fact that the formulae presented in Sec.~A can be used with the substitution $P \rightarrow S = 1 - X - Y$. Hence, the coupled neutrino-plasma upper-hybrid branch has a neutrino solution $N_{+}$ and a plasma solution $N_{-}$, with similar expressions for $\delta Q_{\nu}^{(\pm)}$. The neutrino solution of the upper-hybrid branch is
\begin{equation}
N_{+} \;=\; \frac{1}{K_{\nu}} \left[\; 1 \;+\; \beta\; \frac{(1 - K_{\nu})^{2}}{K_{\nu}\,
(K_{e} - K_{\nu}\,S)} \;+\; \cdots \;\right],
\label{eq:NS_plus}
\end{equation}
and
\begin{equation} 
\delta Q_{\nu}^{(+)}(S) \;\simeq\; -\; \left( \frac{K_{e} - K_{\nu}S}{1 - K_{\nu}} \right) 
\frac{\delta Q_{e}}{\alpha_{e}},
\label{eq:EB_plus}
\end{equation}
while the plasma solution of the upper-hybrid branch is
\begin{equation}
N_{-} \;=\; \frac{1}{K_{e}} \left[\; S \;-\; \beta\; \frac{(K_{e} - S)^{2}}{K_{e}\,
(K_{e} - K_{\nu}\,S)} \;+\; \cdots \;\right],
\label{eq:NS_minus}
\end{equation}
and
\begin{equation} 
\delta Q_{\nu}^{(-)}(S) \;\simeq\; \alpha_{\nu} \left( \frac{S - K_{e}}{K_{e} - K_{\nu}\,S} 
\right) \delta Q_{e}.
\label{eq:EB_minus}
\end{equation}
It is clear that upper-hybrid waves retain their relationship with Langmuir waves as their magnetized analogues \cite{pair_1} even in the presence of collective neutrino-plasma interactions.

\vspace*{0.2in}

\bc
{\sf V. DISCUSSION}
\ec

\vspace*{0.2in}

We now comment on the relations between the perturbed CP-asymmetric densities $\delta Q_{\nu}^{(\pm)}$ and $\delta Q_{e}$ associated with Langmuir and upper-hybrid waves in a magnetized neutrino-plasma medium. The neutrino solution $\delta Q_{\nu}^{(+)}$ associated with the Langmuir and upper-hybrid waves is proportional to $\alpha_{e}^{-1}\delta Q_{e} \gg \delta Q_{e}$. Hence, even a small electron-positron charge separation can lead to a large CP-asymmetric neutrino-antineutrino density $\delta Q_{\nu}^{(+)}$. 

The plasma solution $\delta Q_{\nu}^{(-)}$, on the other hand, is proportional to $\alpha_{\nu}
\,\delta Q_{e} \ll \delta Q_{e}$. Hence, a small electron-positron charge separation leads to a very small CP-asymmetric neutrino-antineutrino density $\delta Q_{\nu}^{(-)}$. We can reverse the argument, however, and say that a CP-asymmetric neutrino-antineutrino density $\delta Q_{\nu}$ can induce a large electron-positron charge separation $\delta Q_{e}^{(-)}$ proportional to $\alpha_{\nu}^{-1}\delta Q_{\nu} \gg \delta Q_{\nu}$.

The present work has thus revealed that collective neutrino-plasma interactions may be quite efficient at breaking the leptonic CP-symmetry of the Early Universe when it is perturbed by space-charge waves (whether a primordial magnetic field exists or not). The role of leptonic asymmetry in Cosmology has received a lot of attention recently within the context of the formation of large-scale structure and the cosmic microwave background anisotropy 
\cite{LSS_1,LSS_2,LSS_3}, and big-bang nucleosynthesis \cite{BBN}. Hence, it is possible that 
in the last era of the Early Universe (just before neutrinos and antineutrinos decoupled from matter), the electron and electron-neutrino lepton number densities $L_{e} = n_{e} - n_{\ov{e}}$ and $L_{\nu} = n_{\nu_{e}} - n_{\ov{\nu}_{e}}$ both experienced large fluctuations as a result of CP-symmetry breaking collective neutrino-plasma interactions. It is perhaps possible that such large fluctuations could be imprinted in the cosmic neutrino background itself \cite{CNB_1,CNB_2}.

\vspace*{0.2in}

\no
{\bf Acknowledgments}

\vspace*{0.1in}

This paper is dedicated to the memory of John M.~Dawson whose work on collective neutrino-plasma interactions inspired one of us (AJB) to pursue this potentially important line of research.

The work presented here was supported by the Vermont EPSCoR Small College Development program and the Vermont Genetics Network.

\vfill\eject

\end{document}